\documentclass[aps,prc,amsfonts]{revtex4}
\usepackage{bm,amsmath,epsfig,dcolumn,slashbox}

\begin{document}

\title{ Formal and Physical ${\bm R}$-matrix parameters }

\author{ Carl~R.~Brune }

\affiliation{Edwards Accelerator Laboratory,
  Department of Physics and Astronomy,
  Ohio University, Athens OH 45701, USA }

\date{Octer 21, 2004}

\begin{abstract}

Notes from 11~October 2004 lecture presented at the Joint Institute
for Nuclear Astrophysics ${\bm R}$-Matrix School at Notre Dame University.

\end{abstract}

%\pacs{24.10.-i,  24.30.-v, 31.15.-p, 02.10.Ud  }
\maketitle

\section{Introduction}

The ${\bm R}$-matrix theory of reactions  has proven over
the course of time to be very useful in nuclear and
atomic physics, both for the fitting of experimental data
and as a tool for theoretical calculations.
In these notes I will explore the relationship between formal
${\bm R}$-matrix parameters and ``physical'' resonance parameters.

\section{How does one define a resonance energy?}
\label{sec:er}

In practice one is likely to encounter many definitions of
``{\em the} resonance energy''~-- e.g. the energy where there is a
peak of an excitation function, the energy when the phase shift
is equal to $\pi/2$ (for elastic scattering), the complex poles
of the scattering matrix,\ldots Many approaches involve the
Breit-Wigner resonance formula:
\begin{equation}
\sigma_{cc'}(E)=\frac{\pi\omega}{k_c^2}\frac{\Gamma^o_c\Gamma^o_{c'}}
  {(E_R-E)^2+(\Gamma^o/2)^2}
\label{eq:bw}
\end{equation}
where $c$($c'$) label the incoming (outgoing) channel, $k_c^2$
is the incoming wavenumber, $\omega=(2J+1)/[(2J_1+1)(2J_2+1)]$ is
the statistical factor, $E$ is the (variable) reaction energy,
$E_R$ is the resonance energy, $\Gamma_c^o$ are the {\em observed} partial
widths of the resonance, and the total {\em observed} width $\Gamma^o$
is given by sum over all channels of the partial widths:
$\Gamma^o = \sum_c \Gamma^o_c$.
It should be noted that the Breit-Wigner formula is a very
general quantum-mechanical result with applications ranging from
atomic physics to particle physics (e.g. the 92-GeV resonance in
$e^+-e^-$ scattering known as the $Z_0$ boson).
In some cases the partial widths may be energy dependent.
From a time-dependent viewpoint the resonant state
decays with time-dependence $\exp(-t/\tau)$, where $\tau$ is the
mean lifetime and $\Gamma^o \tau=\hbar$.
I will describe below one approach for defining resonance energies and
partial widths in the ${\bm R}$-matrix formalism which is
particularly convenient. 
The differences between various prescriptions for defining
resonance energies and partial widths are generally most significant
for broad resonances.
Let me conclude this section with a caution:
since different methods are in use for defining resonance parameters,
it is very important to (1) understand what
conventions are used by others if you use their results and (2)
supply enough information so that others may properly use your results.

\section{Review of ${\bm R}$-Matrix theory}
\label{sec:standardR}

I will begin by reviewing some of the notation and results of
${\bm R}$-matrix theory as described by Lane and Thomas (LT)~\cite{Lan58}
which will be needed later.
The ${\bm R}$ matrix is a function of the energy $E$ and is defined by
\begin{equation}
R_{c'c}=\sum_\lambda \frac{\gamma_{\lambda c'}\gamma_{\lambda c}}
  {E_\lambda -E},
\label{eq:rmatrix}
\end{equation}
where $E_\lambda$ are the level energies,
$\gamma_{\lambda c}$ are the reduced width amplitudes,
$\lambda$ is the level label, and $c$ is the channel label.
I will assume that the numbers of levels and channels
are finite and given by $N_\lambda$ and $N_c$, respectively.
One must also specify the constant(s) $B_c$,
which determine the boundary conditions satisfied by the
underlying eigenfunctions. I will also consider here a set of levels
which all have the same $J^\pi$ value (the presence of additional
levels with other $J^\pi$ values would not impact any results discussed here).

When working in the level-matrix framework, it is convenient
to define column vectors ${\bm \gamma}_c$ in level space from
the $\gamma_{\lambda c}$:
\begin{equation}
{\bm \gamma}_c = \left(
  \begin{array}{l}
    \gamma_{1c} \\
    \gamma_{2c} \\
    \gamma_{3c} \\
    \ldots
  \end{array} \right)
\end{equation}
as well as a diagonal matrix ${\bm e}$ containing the $E_\lambda$:
\begin{equation}
[{\bm e}]_{\lambda \mu} = E_\lambda \delta_{\lambda \mu}.
\end{equation}
The $N_\lambda\times N_\lambda$ level matrix ${\bm A}$ can now
be defined by its inverse:
\begin{eqnarray}
[{\bm A}^{-1}]_{\lambda \mu} &=& (E_\lambda-E)\delta_{\lambda\mu}
  -\sum_c \gamma_{\lambda c}\gamma_{\mu c}(S_c+{\rm i}P_c-B_c) {\rm ,~or} \\
{\bm A}^{-1} &=& {\bm e}- E{\bm 1} -\sum_c {\bm \gamma}_c{\bm \gamma}_c^T
  (S_c + {\rm i}P_c -B_c), \label{eq:ainv}
\end{eqnarray}
where ${\bm 1}$ is the unit matrix and $P_c$ and $S_c$
are the energy-dependent penetration and shift functions.
Note also the notation ${\bm X}^T \equiv {\rm transpose}({\bm X})$
for any matrix ${\bm X}$.
The collision matrix ${\bm U}$ which determines the cross sections
can be calculated from the ${\bm A}$ matrix:
\begin{equation}
[{\bm U}]_{c'c}=\Omega_{c'}\Omega_c\left[
  \delta_{c'c}+2{\rm i}(P_{c'} P_c)^{1/2}
  {\bm \gamma}_{c'}^T {\bm A} {\bm \gamma}_c \right] ,
\label{eq:u_a}
\end{equation}
where the $\Omega_c$ are defined in terms of Coulomb wavefunctions
(see LT, Eq.~III.4.5).

\section{The Single-Level ${\bm R}$-matrix}
\label{sec:one_level}

In order to interpret the meaning of the ${\bm R}$-matrix parameters
$E_\lambda$ and $\gamma_{\lambda c}$ it is very helpful to consider
the case of a single level, i.e. $N_\lambda = 1$. In this case the
${\bm A}$ matrix is simply a number and the cross section is given by
\begin{equation}
\sigma_{cc'}(E)=\frac{\pi\omega}{k_c^2}
  \frac{\Gamma_{\lambda c}\Gamma_{\lambda c'}}
  {(E_\lambda+\Delta_\lambda-E)^2+(\Gamma_\lambda/2)^2},
\label{eq:onelev}
\end{equation}
where
\begin{eqnarray}
\Gamma_{\lambda c} &=& 2\gamma_{\lambda c}^2 P_c, \label{eq:glc}\\
\Delta_\lambda &=& -\sum_c \gamma_{\lambda c}^2 [S_c(E) - B_c] {\rm ,~and} \\
\Gamma_\lambda &=& \sum_c \Gamma_{\lambda c}.
\end{eqnarray}
The partial and total widths $\Gamma_{\lambda c}$ and $\Gamma_\lambda$
are known as {\em formal} widths. This formula for the cross section
is very similar to the Breit-Wigner formula defined above by Eq.~(\ref{eq:bw})
(in fact the single-level ${\bm R}$-matrix formula is sometimes
called the Breit-Wigner formula).
The relationship between Eqs.~(\ref{eq:bw}) and~(\ref{eq:onelev})
is simplest if the $B_c$ are chosen
such that $B_c = S_c(E_\lambda)$. The energy-dependent quantity
$\Delta_\lambda$ can then be expanded in a Taylor series around
$E=E_\lambda$:
\begin{eqnarray}
\Delta_\lambda &=& (E-E_\lambda)\left(
  \frac{\Delta_c}{dE}\right)_{E_\lambda} +...\\
&=& (E_\lambda -E)\sum_c \gamma_{\lambda c}^2
  \left(\frac{dS_c}{dE}\right)_{E_\lambda} +...
\end{eqnarray}
If we replace $\Delta_\lambda$ with the first non-zero term in the
expansion, we have
\begin{equation}
(E_\lambda+\Delta_\lambda-E)^2 \approx  (E_\lambda -E)^2
  \left[1+\sum_c \gamma_{\lambda c}^2 \left(\frac{dS_c}{dE}\right)_{E_\lambda}
  \right]^2 .
\end{equation}
We can now connect the ${\bm R}$-matrix parameters with the
Breit-Wigner resonance parameters. If we identify
\begin{eqnarray}
E_R &=& E_\lambda \quad\quad {\rm and} \label{eq:er_def}\\
\Gamma^o_{\lambda c} &=& \frac{\Gamma_{\lambda c}(E_\lambda)}{1+\sum_c
  \gamma_{\lambda c}^2 \left(\frac{dS_c}{dE}\right)_{E_\lambda}}
  = \frac{2\gamma^2_{\lambda c} P_c(E_\lambda)}{1+\sum_c
  \gamma_{\lambda c}^2 \left(\frac{dS_c}{dE}\right)_{E_\lambda}},
  \label{eq:gam_def}
\end{eqnarray}
then the one-level ${\bm R}$-matrix formula [Eq.~(\ref{eq:onelev})]
and the Breit-Wigner formula [Eq.~(\ref{eq:bw})] are equivalent.
The presence of the penetration factor in Eq.~(\ref{eq:glc}) gives rise
to an energy dependence of the $\Gamma_{\lambda c}$~-- hence the need for
defining $\Gamma_{\lambda c}(E_\lambda)$ and $P_c(E_\lambda)$
in Eq.~(\ref{eq:gam_def}).
Eqs.~(\ref{eq:er_def}) and~(\ref{eq:gam_def}) can be considered as the
definition of the resonance energy and partial widths for the case
of a single-level ${\bm R}$-matrix (i.e. the {\em physical} parameters).
These relationships are also discussed in Sec.~XII.3(a) of LT.
The resonance energy defined in this manner may differ
from the peak of cross section excitation function due to the energy
dependences of the penetration factors and $k_c^2$.
Note also that the difference between formal and
observed partial widths is often
on the order of 30\% -- a large factor compared to typical
experimental uncertainties.

\section{General definition of physical resonance parameters}
\label{sec:gendef}

When there is more than one level, a one-level approximation to the
${\bm R}$ matrix is very good when $E$ is near the ${\bm R}$-matrix
pole energy $E_\lambda$ of a particular level [this approximation
becomes exact when $E\rightarrow E_\lambda$; see Eq.~(\ref{eq:rmatrix})].
Eqs.~(\ref{eq:er_def}) and~(\ref{eq:gam_def}) from the previous section
thus define the physical parameters for that particular level.
However, this approach requires defining the boundary condition
constants such that $B_c=S_c(E_\lambda)$.
The $B_c$ can only be set once (they are after all ``constants'')~--
therefore we only have a simple relationship between ${\bm R}$-matrix
parameters and physical parameters for one of the levels.

One approach to finding the physical parameters for the remaining
levels is to change the $B_c$ to values appropriate
for another physical resonance energy.
The ${\bm R}$-matrix parameters $E_\lambda$ and $\gamma_{\lambda c}$
must then be re-determined~-- either by refitting data or by the
transformation procedure discussed in the next section.
This procedure must be then be repeated one level at a time
until physical parameters have been found for all of the levels.
Note also that this procedure is necessarily iterative -- we want to set
$B_c=S_c(E_\lambda)$ where $E_\lambda$ is a physical resonance energy,
but the physical resonance energy is one of the quantities we are seeking
to determine. Fortunately an iterative approach starting with
the ${\bm R}$-matrix pole energy converges quickly.

\section{How to change the $B_c$}
\label{sec:change_bc}

An interesting feature of ${\bm R}$-matrix theory is that the collision matrix
(i.e. the ${\bm U}$ matrix) is invariant under
changes in the $B_c$, provided that the
$E_\lambda$ and $\gamma_{\lambda c}$ are suitably adjusted.
This result remains true even for the case of finite $N_\lambda$~\cite{Bar72}.
The transformation is most easily described using matrix equations
in level space. Let us consider the transformation $B_c\rightarrow B_c'$,
$E_\lambda\rightarrow E_\lambda'$, and
$\gamma_{\lambda c}\rightarrow\gamma_{\lambda c}'$.
One first constructs the real and symmetric matrix ${\bm C}$ defined by
\begin{equation}
{\bm C}={\bm e}-\sum_c{\bm \gamma}_c{\bm \gamma}_c^T(B_c'-B_c),
\label{eq:cmatrix}
\end{equation}
which is diagonalized by the orthogonal matrix ${\bm K}$
such that ${\bm D}={\bm K}{\bm C}{\bm K}^T$,
with $D_{\lambda\mu}=D_\lambda\delta_{\lambda\mu}$.
Note that any real and symmetric matrix can be diagonalized by
an orthogonal matrix; an orthogonal matrix ${\bm K}$ satisfies
${\bm K}^T={\bm K}^{-1}$ (in other words
${\bm K}^T{\bm K}={\bm K}{\bm K}^T={\bm 1}$).
The necessary transformation of the ${\bm R}$-matrix parameters
is then given by~\cite{Bar72}
\begin{equation}
E_\lambda'=D_\lambda
\label{eq:etrans}
\end{equation}
and
\begin{equation}
{\bm \gamma}_c'={\bm K}{\bm \gamma}_c.
\end{equation}
By using Eq.~(\ref{eq:ainv}) one can show that these transformations leave
${\bm \gamma}_{c'}^T{\bm A}{\bm \gamma}_c$ and hence ${\bm U}$
[see Eq.~(\ref{eq:u_a})] invariant.

The invariance of ${\bm U}$ for changes in $B_c$ has several
important consequences. I will highlight two of them:
(1) mathematically-equivalent cross section fits should be attainable
for any $B_c$ choices and (2) we are free to choose $B_c$ however
we desire. We can exploit this freedom by adopting 
$B_c$ values which are convenient [e.g. $B_c=S_c(E_\lambda)$].
This situation is somewhat analogous to that encountered in electromagnetism:
transformations of the scalar and vector potentials
(known as gauge transformations) can be carried out which leave
the physical electric and magnetic fields invariant.

It is instructive to look at the transformation for the case of a single
level. The matrix ${\bm C}$ consists of only a single dimension,
i.e. a number. The resulting transformation is
\begin{eqnarray}
E_\lambda' &=& E_\lambda - \sum_c \gamma_{\lambda c}^2(B_c'-B_c) \\
\gamma_{\lambda c}' &=& \gamma_{\lambda c}.
\end{eqnarray}
It is straightforward to verify by substitution into the single-level
formula [Eq.~(\ref{eq:onelev})] that the cross section is invariant.

\section{Another Approach to Physical Parameters}

In this section I will describe another mathematically-equivalent
method for extracting physical parameters from ${\bm R}$-matrix parameters.
I begin by defining the real and symmetric matrix ${\bm {\mathcal E}}$:
\begin{equation}
{\bm {\mathcal E}}={\bm e}-\sum_c{\bm \gamma}_c{\bm \gamma}_c^T[S_c(E)-B_c],
\label{eq:ematrix}
\end{equation}
and consider the eigenvalue equation
\begin{equation}
{\bm {\mathcal E}}{\bm a}_i = \tilde{E}_i{\bm a}_i
\label{eq:eigen}
\end{equation}
where $\tilde{E}_i$ is the eigenvalue and
${\bm a}_i$ is the corresponding eigenvector.
Note that ${\bm {\mathcal E}}$
is implicitly dependent upon $\tilde{E}_i$ through $S_c$, so
the eigenvalue problem is nonlinear.
We will assume for convenience that the eigenvectors are normalized
so that ${\bm a}_i^T{\bm a}_i=1$.

The eigenvalues $\tilde{E}_i$ are in fact just the physical
resonance energies. A new set of reduced width parameters is
defined via
\begin{equation}
\tilde{\gamma_{ic}} = {\bm a}_i^T{\bm \gamma}_c.
\end{equation}
The $\tilde{\gamma}_{ic}$ are equal to the ${\bm R}$-matrix reduced
widths when transformed to $B_c=S_c(\tilde{E}_i)$.
The $\tilde{\gamma}_{ic}$ are sometimes called the
``on-resonance'' reduced-width amplitudes.
The observed (physical) partial widths are thus given by
\begin{equation}
\Gamma_{i c}^o=\frac{2P_c(\tilde{E}_i)\tilde{\gamma}_{ic}^2}
  {1+\sum_c\tilde{\gamma}_{ic}^2
\left(\frac{dS_c}{dE}\right)_{\tilde{E}_i}}.
\label{eq:Gamma}
\end{equation}
One advantage of this formulation is that the solution of the
eigenvalue problem gives immediately all of the physical resonance
energies and partial widths.

However one defines the physical resonance parameters, it is
desirable that they be independent of the boundary condition
constants and channel radii of ${\bm R}$-matrix theory.
The above approach is completely independent of the $B_c$ and
approximately independent of the channel radii.
Additional details and the justification for this approach are
given by Brune~\cite{Bru02}. The ``on-resonance'' reduced-width
amplitudes defined here are also disused by Barker~\cite{Bar71}
and by Angulo and Descouvemont~\cite{Ang00}.

I would like to mention in passing another similar eigenvalue equation.
If we had instead defined
\begin{equation}
{\bm {\mathcal E}}={\bm e}-\sum_c{\bm \gamma}_c{\bm \gamma}_c^T
[S_c(E)+{\rm i}P_c(E)-B_c],
\end{equation}
we would have the eigenvalue equation
introduced by Hale, Brown, and Jarmie~\cite{Hal87}.
The eigenvalues of this equation are complex and correspond
to complex poles of the ${\bm A}$ and ${\bm U}$ matrices
(the latter is also known as the scattering matrix).
As discussed in Refs.~\cite{Hal87,Til02} this approach to defining
resonance energies and widths may be theoretically preferable.
However, there are significant complexities associated with
calculating the Coulomb wavefunctions for complex momenta.
We will not discuss this approach further here.

\section{Converting Physical parameters to ${\bm R}$-matrix parameters}

Next we will investigate how to translate physical parameters into
${\bm R}$-matrix parameters~\cite{Bru02}.
It is often desirable to incorporate
information from a level compilation or shell model calculation
into an ${\bm R}$-matrix parameterization. Values for $\Gamma_{ic}^o$
must first be converted into reduced widths $\tilde{\gamma}_{ic}$ using
Eq.~(\ref{eq:Gamma}). This procedure requires the straightforward 
solution of a system of linear equations for $\tilde{\gamma}_{ic}^2$;
the only complication is that
the {\em signs} of reduced-width amplitudes cannot be determined.
Relative signs can be determined via an ${\bm R}$-matrix
calculation (which automatically includes interference effects between levels)
and a comparison to experimental data.

With the physical resonance energies $\tilde{E}_i$ and
``on-resonance'' reduced-width amplitudes $\tilde{\gamma}_{ic}$
in hand we can determine ${\bm R}$-matrix parameters.
We proceed by defining the matrices ${\bm M}$ and ${\bm N}$:
\begin{equation}
[{\bm M}]_{ij}\equiv\left\{
\begin{array}{ll} 1 & i=j \\ -\sum_c\tilde{\gamma}_{ic}\tilde{\gamma}_{jc}
\frac{S_{ic}-S_{jc}}{\tilde{E}_i-\tilde{E}_j} & i\neq j
\end{array} \right.
\label{eq:mmatrix}
\end{equation}
and
\begin{equation}
\label{eq:nmatrix}
[{\bm N}]_{ij} \equiv \left\{
\begin{array}{ll}\tilde{E}_i+\sum_c \tilde{\gamma}_{ic}^2(S_{ic}-B_c) & i=j \\
\sum_c \tilde{\gamma}_{ic}\tilde{\gamma}_{jc}
\left(\frac{\tilde{E}_iS_{jc}-\tilde{E}_jS_{ic}}{\tilde{E}_i-\tilde{E}_j}
-B_c \right) & i\neq j \end{array} \right. ,
\end{equation}
where $S_{ic}$ denotes $S_c(\tilde{E}_i)$.
Note that the construction of ${\bm N}$ requires the adoption
of specific $B_c$ values.
The eigenvalue equation
\begin{equation}
{\bm N} {\bm b}_\lambda = E_\lambda {\bm M} {\bm b}_\lambda.
\label{eq:master}
\end{equation}
now holds the key for transforming from the
$\tilde{E}_i$-$\tilde{\gamma}_{ic}$ representation to the
standard ${\bm R}$-matrix parameters $E_\lambda$ and $\gamma_{\lambda c}$.
The eigenvalues are the ${\bm R}$-matrix pole energies $E_\lambda$
and the eigenvectors ${\bm b}_\lambda$ yield the ${\bm R}$-matrix
reduced-width amplitudes:
\begin{equation}
\gamma_{\lambda c} = {\bm b}_\lambda^T \tilde{\bm \gamma}_c.
\label{eq:gtrans}
\end{equation}
The matrices ${\bm M}$ and ${\bm N}$ are
real, symmetric, and energy-independent.
The solution to this type of eigenvalue problem is discussed in
Sec.~8.7.2 of Ref.~\cite{Gol96}; I have utilized the
LAPACK~\cite{lapack} routine {\sc dsygv}.

The eigenvectors ${\bm b}_\lambda$ can be
arranged into a square matrix ${\bm b}$ which simultaneously
diagonalizes ${\bm M}$ and ${\bm N}$:
\begin{equation}
{\bm b}^T {\bm M} {\bm b} = {\bm 1}
\label{eq:mb_diag}
\end{equation}
and
\begin{equation}
{\bm b}^T {\bm N} {\bm b} = {\bm e}.
\end{equation}
We can also re-write Eq.~(\ref{eq:gtrans}) as
\begin{equation}
{\bm \gamma}_c = {\bm b}^T \tilde{\bm \gamma}_c .
\label{eq:conv_gamma}
\end{equation}

\section{Working Directly With Physical Parameters}

It turns to be possible to carry out ${\bm R}$-matrix calculations
directly from the physical parameters~\cite{Bru02}.
This result can be achieved by defining a new (alternative)
level matrix $\tilde{\bm A}$ from the physical resonance energies
$\tilde{E}_i$ and ``on-resonance'' reduced-width amplitudes
$\tilde{\gamma}_{ic}$:
\begin{eqnarray}
\label{eq:alta_inv}
[\tilde{\bm A}^{-1}]_{ij} &=& (\tilde{E}_i-E)\delta_{ij}-\sum_c
\tilde{\gamma}_{ic} \tilde{\gamma}_{jc} (S_c+{\rm i}P_c) \\ \nonumber
&+& \sum_c \left\{ \begin{array}{ll}
\tilde{\gamma}_{ic}^2 S_{ic} & i=j \\
\tilde{\gamma}_{ic} \tilde{\gamma}_{jc}
\frac{S_{ic}(E-\tilde{E}_j) - S_{jc}(E-\tilde{E}_i)}
{\tilde{E}_i-\tilde{E}_j} & i\neq j \end{array} \right. .
\end{eqnarray}
This matrix is then used with the $\tilde{\bm \gamma}_{c}$ to calculate the
the collision matrix ${\bm U}$:
\begin{equation}
U_{c'c}=\Omega_{c'}\Omega_c\left[ \delta_{c'c}+2{\rm i}(P_{c'} P_c)^{1/2}
\tilde{\bm \gamma}_{c'}^T \tilde{\bm A} \tilde{\bm \gamma}_c \right] .
\label{eq:u_alta}
\end{equation}
This method of calculating ${\bm U}$ is mathematically equivalent to
first transforming the physical $\tilde{E}_i$-$\tilde{\gamma}_{ic}$
parameters to ${\bm R}$-matrix parameters $E_\lambda$-$\gamma_{\lambda c}$
and then calculating ${\bm U}$ using standard ${\bm R}$-matrix equations.
Note that in this approach the boundary-condition constants $B_c$
never need to be defined at all. One application of this approach is
the possibility of using the physical parameters directly as fit parameters.

\begin{table}[tb]
\caption{Standard ${\bm R}$-matrix parameters from Table~III
of Ref.~\protect\cite{Azu94} which describe $J^\pi=1^-$ reactions
in the ${}^{16}{\rm O}$ system, and the physical parameters
derived from them. 
The channel radius is 6.5~fm and the boundary condition constant is
chosen according to  $B_c=S_c(E_1)$.
The $\beta$-decay feeding amplitudes
${\mathcal B}_\lambda$ are equivalent to the quantities
$A_{\lambda 1} \gamma_{\lambda 1}^{-1} N_\alpha^{-1/2}$ of
Ref.~\cite{Azu94}.}
\begin{ruledtabular}
\begin{tabular}{lddd}
$\lambda$ & \multicolumn{1}{c}{1} & \multicolumn{1}{c}{2} &
  \multicolumn{1}{c}{3} \\ \hline
$E_\lambda$ (MeV) & -0.0451 & 2.845 & 11.71 \\
$\gamma_{\lambda\alpha}$ (MeV$^{1/2}$) & 0.0793 & 0.330 & 1.017 \\
$\gamma_{\lambda\gamma}$ (MeV$^{-1}$) &
  \multicolumn{1}{c}{$8.76\times 10^{-6}$} &
  \multicolumn{1}{c}{$-2.44\times 10^{-6}$} &
  \multicolumn{1}{c}{$-2.82\times 10^{-6}$} \\
${\mathcal B}_\lambda$ & 1.194 &  0.558 & -0.629 \\ \hline
$\tilde{E}_\lambda$ (MeV) & -0.0451 & 2.400 & 8.00 \\
$\tilde{\gamma}_{\lambda\alpha}$ (MeV$^{1/2}$) & 0.0793 & 0.471 & 0.912 \\
$\tilde{\gamma}_{\lambda\gamma}$ (MeV$^{-1}$) &
  \multicolumn{1}{c}{$8.76\times 10^{-6}$} &
  \multicolumn{1}{c}{$-3.20 \times 10^{-6}$} &
  \multicolumn{1}{c}{$-2.50\times 10^{-6}$} \\
$\tilde{\mathcal B}_\lambda$ & 1.194 & 0.408 & -0.781 \\
$1+\tilde{\gamma}_{\lambda\alpha}^2 \frac{dS_\alpha}{dE}(\tilde{E}_\lambda)$ &
 1.0050 & 1.2034 & 1.0220 \\
$\Gamma_\alpha^o$ (MeV) & - & 0.359 & 9.041  \\
$\Gamma_\gamma^o$ (meV) & 55.1 & 8.64 & 54.2
\end{tabular}
\end{ruledtabular}
\label{tab:params}
\end{table}

\section{An Example}

At this point it is instructive to introduce a concrete example.
The top section of Table~\ref{tab:params} shows
a simple well-documented set of standard
${\bm R}$-matrix parameters taken from Azuma {\em et al.}~\cite{Azu94}.
This analysis concerns the low-lying $J^\pi=1^-$ states of $^{16}{\rm O}$
reacting through channels $^{12}{\rm C}+\alpha$ ($c=\alpha$, $l_\alpha=1$)
and $^{16}{\rm O}+\gamma$ ($c=\gamma$, $E1$ multipolarity).
The zero of the energy scale is at the $^{12}{\rm C}+\alpha$ threshold.
Note that the gamma-ray channel is treated perturbatively
and does enter in the calculation of the ${\bm A}$ matrix.

The bottom section of Table~\ref{tab:params} shows
physical resonance energies $\tilde{E}_i$ and the ``on-resonance''
reduced-width amplitudes $\tilde{\gamma}_{\lambda c}$ and
$\beta$-decay feeding factors ${\mathcal B}_i$.
In addition the physical partial widths $\Gamma_{ic}^o$ are
given along with values of
$[1+\tilde{\gamma}_{i\alpha}^2 \frac{dS_\alpha}{dE}(\tilde{E}_i)]$
needed to related formal and physical widths.

This example is also useful for demonstrating some of the other
concepts discussed here. For example one can convert from the
$\tilde{E}_i$-$\tilde{\gamma}_{ic}$ representation back to the
standard ${\bm R}$-matrix $E_\lambda$-$\gamma_{\lambda c}$ representation.
In Table~\ref{tab:b_elements} we show the elements of the
transformation matrix ${\bm b}$
[see Eqs.~(\ref{eq:mb_diag}-\ref{eq:conv_gamma})] needed
to carry the transformation back to standard ${\bm R}$-matrix parameters.
One can also verify that the various approaches for calculating
the collision matrix ${\bm U}$
[e.g. Eqs.~(\ref{eq:u_a}) and~(\ref{eq:u_alta})] yield the
same result.

\begin{table}[tb]
\caption{Elements $[{\bm b}]_{ij}$ of the transformation matrix
corresponding to the parameters of Table~\ref{tab:params}.}
\begin{ruledtabular}
\begin{tabular}{cddd}
\backslashbox{$i$}{$j$}  & \multicolumn{1}{c}{1} & \multicolumn{1}{c}{2} &
  \multicolumn{1}{c}{3}      \\ \hline
1  & 1.000 & 0.0373  & 0.0446 \\
2  & 0.000 & 0.9781  & 0.2281 \\
3  & 0.000 & -0.1466 & 0.9933
\end{tabular}
\end{ruledtabular}
\label{tab:b_elements}
\end{table}

\section{Summary}

Several aspects of the relationship between ${\bm R}$-matrix
parameters and physical resonance parameters have been discussed.
First it is necessary to establish how physical resonance energies
and partial widths will be defined. By comparing the one-level
${\bm R}$-matrix formula for the cross section to the Breit-Wigner
formula it is possible to define a relationship between
${\bm R}$-matrix parameters and physical resonance parameters.
For the multi-level case the extraction of physical parameters is
more complicated but can be achieved by the refitting data with
different $B_c$ values, carrying out $B_c$ transformations, or
by solving an eigenvalue equation. We next investigated the reverse
procedure, i.e. transforming physical resonance parameters into
${\bm R}$-matrix parameters, which is useful for incorporating information
from other sources into an ${\bm R}$-matrix fit.
A method for calculating observables directly from the ``on-resonance''
${\bm R}$-matrix parameters was then discussed.
Finally we presented an example case which is useful for
demonstrating the ideas which have been presented here.

\end{document}